\documentclass{emulateapj}
\slugcomment{Accepted by  {\it The Astrophysical Journal}}
\def\farcm{\hbox{$.\mkern-4mu^\prime$}}
\def\farcs{\hbox{$~\mkern-4mu^{\prime\prime}$}}

\def\la{\mathrel{\hbox{\rlap{\hbox{\lower4pt\hbox{$\sim$}}}\hbox{$<$}}}}
\def\ga{\mathrel{\hbox{\rlap{\hbox{\lower4pt\hbox{$\sim$}}}\hbox{$>$}}}}

\shortauthors{Schenck}
\shorttitle{SNR B0049-73.6}

\begin{document}

\title{A Deep Chandra Observation of Oxygen-Rich Supernova Remnant B0049--73.6 in the Small Magellanic Cloud}

\author{Andrew Schenck\altaffilmark{1}, Sangwook Park\altaffilmark{1}, David N. Burrows\altaffilmark{2}, John P. Hughes\altaffilmark{3}, Jae-Joon Lee\altaffilmark{4}, and Koji Mori\altaffilmark{5}}

\altaffiltext{1}{Box 19059, Department of Physics, University of Texas at Arlington, Arlington, TX 76019; andrew.schenck@mavs.uta.edu}
\altaffiltext{2}{Department of Astronomy and Astrophysics, Pennsylvania State University, 525 Davey Laboratory, University Park, PA 16802}
\altaffiltext{3}{Department of Physics and Astronomy, Rutgers University, 136 Frelinghuysen Road, Piscataway, NJ 08854-8019}
\altaffiltext{4}{Korea Astronomy and Space Science Institute, Daejeon, 305-348, Korea}
\altaffiltext{5}{Department of Applied Physics, University of Miyazaki, 1-1 Gakuen Kibana-dai Nishi, Miyazaki, 889-2192, Japan}

\begin{abstract}

We report on the initial results from our deep {\it Chandra} observation (450 ks) of O-rich supernova remnant (SNR) B0049--73.6 in the Small Magellanic Cloud. We detect small metal-rich ejecta features extending out to the outermost boundary of B0049--73.6, which were not seen in the previous data with a shorter exposure. The central nebula is dominated by emission from reverse-shocked ejecta material enriched in O, Ne, Mg, and Si. O-rich ejecta distribution is relatively smooth throughout the central nebula. In contrast the Si-rich material is highly structured. These results suggest that B0049--73.6 was produced by an asymmetric core-collapse explosion of a massive star. The estimated abundance ratios among these ejecta elements are in plausible agreement with the nucleosynthesis products from the explosion of a 13--15$M_{\odot}$ progenitor. The central ring-like (in projection) ejecta nebula extends to $\sim$9 pc from the SNR center. This suggests that the contact discontinuity may be located at a further distance from the SNR center than the previous estimate. We estimate the Sedov age of $\sim$17000 yr and an explosion energy of {\it $E_0$} $\sim1.7 \times~10^{51}$ erg for B0049--73.6. We place a stringent upper limit on the 2-7 keV band luminosity of $L_{X}\sim$ 8.5 $\times  10^{31}$ erg s$^{-1}$ for the embedded compact stellar remnant at the center of B0049--73.6.

\end{abstract}

\keywords {ISM: individual objects (B0049--73.6) --- ISM: supernova remnants --- X-rays: ISM}

\section {\label{sec:intro} INTRODUCTION}

Core-collapse (CC) explosions of massive stars ($M >8M_{\odot}$) account for $\sim$3/4 of all supernovae (SNe). These CC SNe directly impact the chemical evolution of the interstellar medium (ISM) surrounding them and, subsequently, the star-formation history of the host galaxy. X-ray observations of CC supernova remnants (SNRs) are useful to study some fundamental natures of CC SNe and their evolution such as the metal composition in the progenitor star, the explosive nucleosynthesis, and the chemical composition of the ISM into which the blast wave is expanding.

B0049--73.6 is a relatively old SNR ($\sim$14,000 yr, Hendrick et al. 2005, H05 hereafter) in the Small Magellanic Cloud (SMC). Using the 50 ks {\it Chandra} observation H05 were able to resolve a bright central nebula surrounded by a faint outer shell. H05 found that the central nebula showed enhanced line emission from highly ionized H-like O and He- and H-like Ne ions. Atomic line emission from He-like Mg and Si ions also appeared enhanced. H05 concluded that the central nebula was metal-rich, particularly in O and Ne, which represents chemically-enriched stellar debris. Based on this ejecta composition H05 identified B0049--73.6 as the third member of the O-rich SNRs in the SMC (the other two being SNRs E0102--7219 [Dopita \& Tuohy 1984] and 0103--72.6 [Park et al. 2003]) and thus a CC SNR. The central ejecta nebula shows a distinct ring-like morphology (in projection) and is complicated in nature with apparent radial and azimuthal sub-structures. Assuming a common $\sim$$15M_{\odot}$ red supergiant (RSG) progenitor for CC SNe H05 performed simple 1D hydrodynamic numerical simulations of SNRs to study the observed structure of B0049--73.6. They found that a bright X-ray ejecta shell would form just interior to the contact discontinuity (CD) that would have a radius of $\sim$10 pc at the age of B0049--73.6. The observed bright O-rich ring in B0049--73.6 has a significantly smaller size (6 pc in radius) than that of the simulated CD. Thus H05 proposed an alternative scenario of an Fe-Ni bubble for the origin of the observed central ring-like nebula. However, its true origin remains elusive.

Here we report on the initial results from our deep $\sim$450 ks {\it Chandra} observation of B0049--73.6. Based on these new {\it Chandra} data, which are $\sim$8 times deeper than those used by H05, we estimate that the true extent of the central ejecta nebula is larger than the size of the bright O-rich ring, more consistent with H05's 1D hydrodynamical model. We detect several new emission features that suggest a highly asymmetrical distribution of metal-rich ejecta in B0049--73.6. In Section 2 we describe our observations and data reduction. In Section 3 we present the results from our imaging and spectral data analyses. A discussion and conclusions are presented in Section 4.

\section{\label{sec:obs} OBSERVATIONS \& DATA REDUCTION}

We performed our {\it Chandra} observations of B0049--73.6, which comprise 17 ObsIDs, between 2010 April 14 and 2010 June 2. We chose the ACIS-S3 instrument with the SNR centered on the aim point to utilize the high sensitivity of the S3 chip in the soft X-ray band ($E$ $\la$ 1 keV). Data were gathered in Very Faint mode and processed with CIAO version 4.5 via the {\it chandra$\_$repro} script with CALDB version 4.5.5, which includes corrections for the charge transfer inefficiency and recent change of the contamination rate in the optical blocking filter, which affected the quantum efficiency of the ACIS by an additional $\sim$15$\%$ since mid-2009. Filtering was done for the standard ASCA grades (0,2,3,4,6) and for photon energies between 0.3 and 3.0 keV, beyond which the SNR emission is negligible. The overall light curve showed no significant flaring background. We performed this data reduction on each individual observation. After the data reduction the total effective exposure is $\sim$441 ks. 

\section{\label{sec:result} ANALYSIS \& RESULTS}

\subsection{IMAGING}
\begin{figure*}[]
\figurenum{1}
\centerline{\includegraphics[angle=0,width=\textwidth]{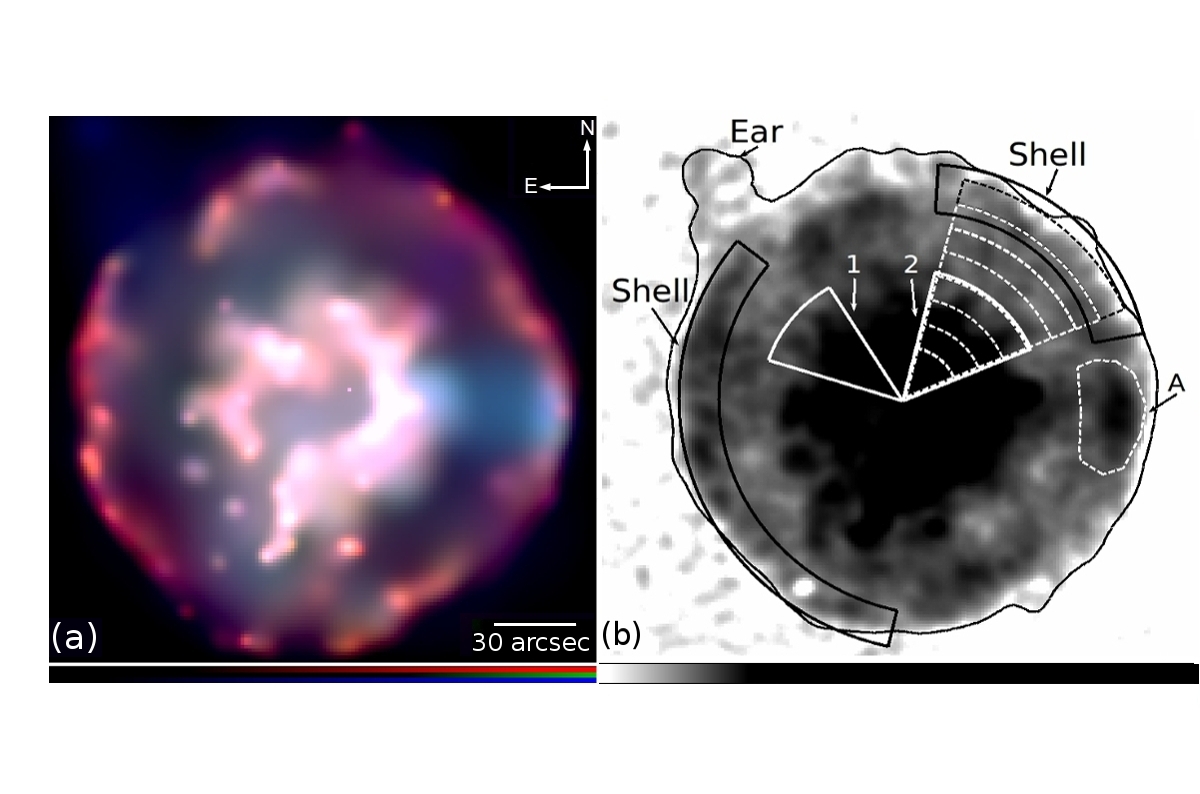}}
\figcaption[]{{(a) A three-color image of B0049--73.6. Color codes are red: 300--700 eV, green: 700--1100 eV, and blue: 1100--3000 eV. Each color image has been binned by 2$\times$2 pixels and then adaptively smoothed. The red and blue bands are on a square-root scale while the green band is on a log scale. (b) A grey-scale broadband image of B0049-73.6. The radial width of the dashed regions is $\sim$2.3 pc (8$\farcs$) except for the innermost region for which it is $\sim$4.6 pc (16$\farcs$). The central ring is saturated in black to emphasize faint outer regions.  The overlaid contour marks the outermost boundary of B0049--73.6 based on the broadband image.}\label{fig:fig1}}
\end{figure*}

Figure~\ref{fig:fig1} shows an X-ray three-color image of B0049--73.6. To create this image, we combined all individual observations. We ran the CIAO tool {\it wavdetect} on the combined data to detect point sources. Although all detected point sources are faint (the combined point source flux makes up $\sim$2$\%$ of the total SNR flux) we removed all of them before further data analysis. We then adaptively smoothed each sub-band image. The small-scale ($\sim$several arcseconds) clumpy features in this image are likely an artifact of the adaptive smoothing. The main outermost boundary shows a nearly circular shell-like morphology and is spectrally soft (red). This outer shell was identified as the swept-up ISM by H05, and our spectral analysis confirms it (see Section 3.2). The radius of the outermost boundary is 1$\farcm$2 from the SNR center. We detect a faint emission feature in the northeast extending beyond the main outermost shell (the ``Ear'' in Figure 1b) which was not detected in the previous data. In contrast to the outer shell, the central nebulosity shows various colors throughout, which indicate a significant contribution from higher energy photons, probably from line emission of various elemental species. This complex color structure of the central nebula is generally consistent with the metal-rich ejecta origin suggested by  H05. The central ring-like nebulosity peaks in brightness at $\sim$20$\farcs$ from the SNR center. This intensity peak corresponds to the O-rich ring identified by H05. Based on our data with a deep exposure, we detect the diffuse emission defining the main outer boundary of this central ejecta nebula, which extends beyond the broadband intensity peak out to $\sim$40$\farcs$ from the SNR center. The central nebula shows asymmetric emission features including an elongated feature in the southern part and various areas of deformation primarily in the east. It is remarkable that a spectrally-hard (blue) emission feature in the western part of the central ejecta nebula extends out to the western outermost shell (Region A in Figure 1b). 

\begin{figure*}[]
\figurenum{2}
\centerline{\includegraphics[angle=0,width=\textwidth]{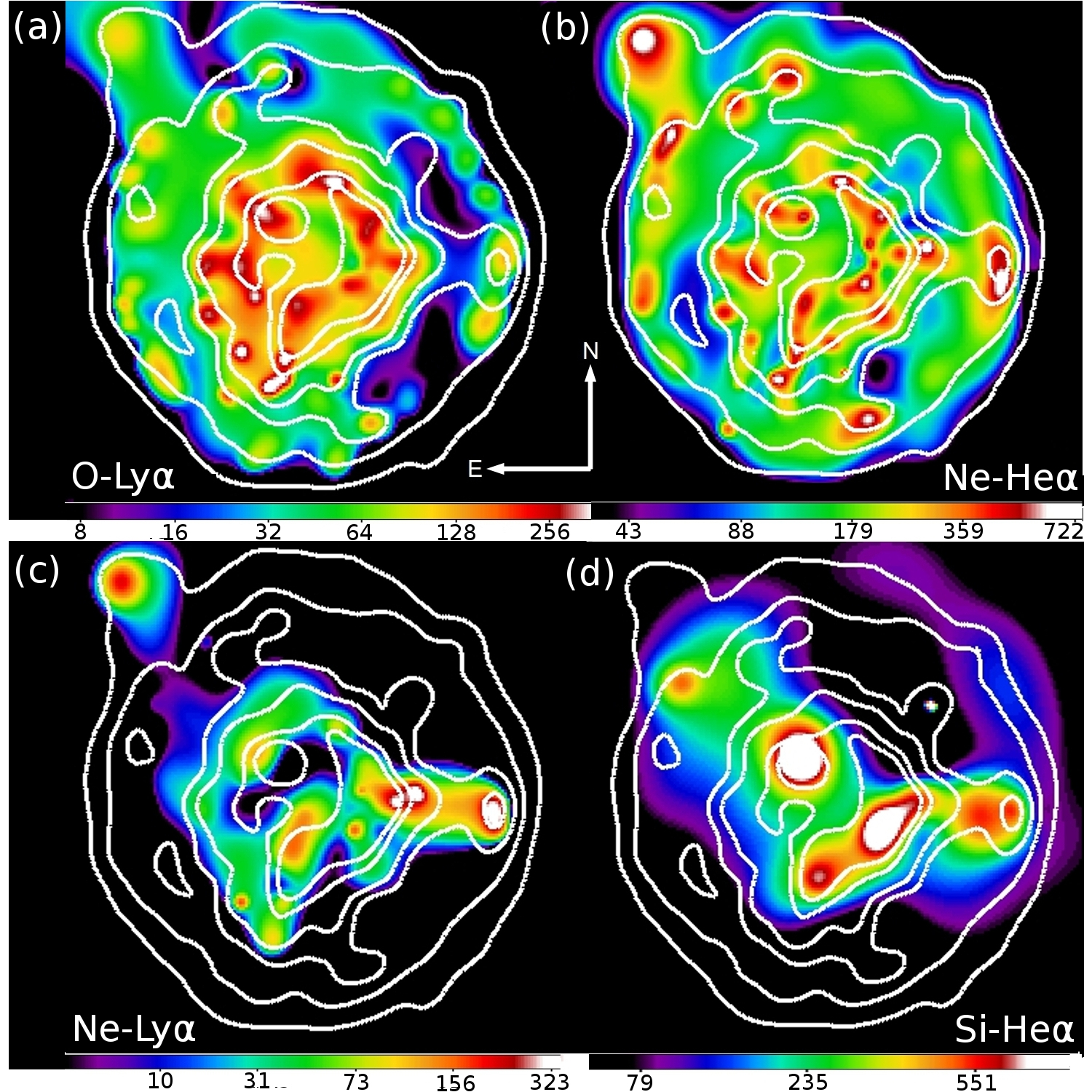}}
\figcaption[]{{Line equivalent-width images of various elemental species for B0049--73.6. The false-color scales are in units of eV. Broadband image contours of the SNR are overlaid.}\label{fig:fig2}}
\end{figure*}

Figure~\ref{fig:fig2} shows {\it equivalent-width} images (EWI) for O-Ly$\alpha$ ({\it E} $\sim$ 0.65 keV), Ne-He$\alpha$ ({\it E} $\sim$ 0.9 keV), Ne-Ly$\alpha$ ({\it E} $\sim$ 1.02 keV), and Si-He$\alpha$ ({\it E} $\sim$ 1.84 keV) line emission. To create these EWI maps we followed the method found in literature \citep[e.g.,][]{hwang00,park02}. We selected emission line bands: O-Ly$\alpha$ = 620--650, Ne-He$\alpha$ = 850--950, Ne-Ly$\alpha$ = 1000--1100, and Si-He$\alpha$ = 1700--1900 eV. Images in these bands were binned by 2$\times$2 pixels and then adaptively smoothed prior to EW calculations. The underlying continuum was calculated by logarithmically interpolating the fluxes from the higher and lower continuum images of narrow energy bands nearby the line to the line center energy. The estimated continuum flux was integrated over the line band and subtracted from the line emission flux. The continuum-subtracted line intensity was then divided by the estimated continuum flux on a pixel-by-pixel basis to generate the EWIs for each line. In order to avoid noise in the EW maps caused by pixels with poor photon statistics, we set the EW values to zero on pixels where the calculated continuum flux is $\la10\%$ of the total flux. Small scale variations in Figure 2 ($\la$ several arcseconds which correspond to typical smoothing scales) are likely statistical artifacts. We note that we use these EW images only for a qualitative guide to track line-enhanced/suppressed areas for an efficient regional spectral analysis. We do not attempt any quantitative interpretations directly from the EW images. We extracted several  characteristic regions based on our EW images (Region A, Ear, and Regions 1 and 2 in Figure 1), and found that actual X-ray spectra from regions of our interest are consistent with the features implied by the EW images (see Section 4.1).

O-Ly$\alpha$ EW is enhanced throughout the entire central ring-like nebulosity. While Ne-He$\alpha$ EW distribution is generally similar to that of O-Ly$\alpha$, several enhanced regions extend beyond the central ring toward the outermost boundary in the west (Region A in Figure 1b), northeast, and south. Ne-He$\alpha$ EW is also enhanced in the Ear. Ne-Ly$\alpha$ enhancement is concentrated in the west (Region A) and in the Ear. Si-He$\alpha$ EW is enhanced mainly in three distinct regions, two within the central nebulosity, and another in Region A. Except for several small EW-enhanced features extending to the outer boundary, these EWIs are generally consistent with the previously-suggested metal-rich ejecta origin for the central ring-like nebulosity. On the other hand, the outer shell shows no enhancement in any EWI except for some small features discussed above (e.g. Region A). These low EWs in the outer shell are in general agreement with its previously-identified origin as the swept-up ISM.

\subsection{SPECTROSCOPY}
We extracted the spectrum from the Shell region shown in Figure 1b. We extracted this spectrum from each individual observation and then combined them using the CIAO script {\it combine$\_$spectra}. Some small parts of the outer shell show various colors while most of the shell is nearly pure red (Figure 1a). These small regions were excluded from our study of the swept-up ISM. The entire Shell region contained $\sim$27,000 counts ($\sim$17,200 counts from the southeastern region and $\sim$9,800 counts from the northwestern region). We binned the spectrum to contain a minimum of 20 counts per energy channel. We fit the Shell spectrum using a nonequilibrium ionization (NEI) plane-parallel shock model \citep{bork01} with two foreground absorption column components, one for the Galactic ($N_{H,Gal}$) and the other for the SMC ($N_{H,SMC}$). We used NEI version 2.0 (in xspec) associated with ATOMDB \citep{smith01} which was augmented to include inner shell lines and updated Fe-L lines (see Badenes et al. 2006). We performed the background subtraction using the spectrum extracted from source-free regions outside of the SNR. We fixed $N_{H,Gal}$ at $4.5 \times 10^{20}$ cm$^{-2}$ for the direction toward B0049--73.6 \citep{DL90}. We fixed abundances in $N_{H,SMC}$ at values by Russell $\&$ Dopita (1992, RD92 hereafter). We varied electron temperature ($kT$, where {\it k} is the Boltzmann Constant), ionization timescale ($n_{e}t$, where $n_{e}$ is the post-shock electron density and $t$ is the time since shock), and normalization in the plane-shock model. We initially fixed all elemental abundances in the plane-shock model at the values found in RD92. The fit was statistically rejected ($\chi_{\nu}^2=3.3$). Residuals from the best-fit model were significant around O and Ne line energies. We refit the data with O and Ne abundances varied. This fit improved but it was still statistically unacceptable ($\chi_{\nu}^2=2.0$). We then varied each additional elemental abundance (Mg, Si, and Fe) one at a time to improve the fit. We found that it was necessary to vary O, Ne, Mg, and Fe to obtain the best-fit model ($\chi_{\nu}^2=1.4$, Figure 3a). This $\chi_{\nu}^2$ value is still too large for a statistically acceptable fit, which we attribute to incompleteness in the NEI model details and systematic errors in the instrument calibration. We note that varying the Si abundance did not make a statistically significant improvement in the fit based on an F-test (F-probability = 0.4). It is notable that the best-fit abundances for the fitted elements are significantly lower (by a factor of $\sim$3--4) than those by RD92. This low abundance is unlikely an artifact caused by underlying synchrotron emission because the observed X-ray spectrum is very soft and there is no observational evidence of thin filaments whose X-ray spectrum is dominated by a spectrally-hard continuum in this relatively old SNR. Based on our inferred abundances and applicability of a single-shock model, we conclude that the Shell region is the swept-up ISM with very low metal abundances. Our best-fit model parameters for the Shell region are listed in Table 1. 

The Region A's spectrum contains $\sim$8000 counts, and shows distinct features compared to the swept-up ISM (Figure 3a). A broad line complex at $E=$ 0.8--1.1 keV along with evident Mg and Si lines are present. This spectrum cannot be fit by a single shock model with abundances that we estimated for the Shell region ($\chi_{\nu}^2=3.2$). These indicate the presence of an additional shock component superposed with the Shell emission. Thus, we used a two-component NEI shock model to fit this spectrum, one for the underlying Shell spectrum and the other responsible for the distinctive component in Region A. We fixed $N_{H,SMC}$ at the Shell value. We fixed all model parameters (except for normalization) of the underlying swept-up ISM component at the values for the Shell region. For the second shock component we initially allowed $kT$, $n_{e}t$, and normalization to vary and fixed the elemental abundances at the Shell values. The fit was not statistically acceptable ($\chi_{\nu}^2=3.0$) because the model was not able to reproduce emission lines from various elements. We then thawed elemental abundances for the second component to improve the fit. The best-fit model required us to fit O, Ne, Mg, and Si ($\chi_{\nu}^2=1.4$). This somewhat large $\chi_{\nu}^2$ value suffers from the same model and detector issues as stated in the ISM discussion, and thus we consider them satisfactory. The Ne and Si abundances are significantly enhanced compared to those of the Shell region: by $\sim$an order of magnitude for Ne and factor of $\sim$2 for Si. O and Mg abundances are within 90\% error of those for the Shell region. The best-fit model parameters for Region A are listed in Table 1. 

We fit Regions 1 ($\sim$12,500 counts) and 2 ($\sim$16,800 counts) spectra (Figure 1b) using a one-component plane-shock model (Figure 3b). Initially we varied $kT$, $n_{e}t$, and normalization while fixing $N_{H,Gal}$, $N_{H,SMC}$, and all elemental abundances at the Shell values. The fit was statistically rejected for both regions ($\chi_{\nu}^2\ga4.0$). We then varied abundances for elements for which emission line features are associated with significant residuals from the fitted model. We found that it was necessary to vary O, Ne, Mg, Si, and Fe abundances in Region 1 to obtain a satisfactory fit ($\chi_{\nu}^2=1.4$). It was necessary to vary O, Ne, Mg and Si abundances for Region 2 ($\chi_{\nu}^2=1.6$). These large $\chi_{\nu}^2$ values suffer from the same model and detector issues as stated in the discussion of the ISM, and thus we consider them satisfactory. The overall abundances for Regions 1 and 2 are significantly enhanced compared to the Shell abundances (by a factor of $\sim$3--6 times). Although our single-shock model fits for Regions 1 and 2 are statistically satisfactory, they may have not adequately accounted for the superposed outer shell emission. To test this geometrical effect, we re-fit these spectra with a two-component shock model, adding a plane shock model component for the superposed outer shell spectrum. For the outer shell component we fixed all parameters at the Shell values except for normalization. The addition of the outer Shell component did not change the results because the contribution from the Shell is small ($\sim$10--20$\%$ of the total flux). Thus we used our results from the one-shock model fits as summarized in Table 1 for the discussion of Regions 1 and 2 in the following sections.

The overall central ejecta nebula appears to extend beyond the intensity peak of the central ring (Figure 1a). To verify this extent of the central ejecta feature we investigate radial distributions of metal abundances based on spectra from 8 radial regions (dashed regions in Figure 1b.) We choose this azimuthal sector because it shows well-defined structures of both the central ejecta nebula and the outer shell along the radius of the SNR. The extracted spectrum from each of these eight radial sectors contains $\sim$2500--4500 counts. We initially fit each regional spectrum with the one-component plane-shock model with all elemental abundances fixed at the Shell values. We allowed electron temperature, ionization timescale, and normalization to vary while fixing $N_{H,SMC}$ to the Shell value. This single shock model fit is satisfactory in the outer regions at $r\ga 40 \farcs$ ($\chi_{\nu}^2\la 1.5$). For inner regions at $r\la 40 \farcs$ the single shock model fits are not statistically acceptable ($\chi_{\nu}^2 \ga 2.0$, Figure 4b), probably because of a significant contribution from central ejecta emission. Thus, we used the two-component shock model for the inner regions at $r \la 40 \farcs$, one for the ejecta component and one for the superposed outer shell component. The parameters (except for normalization) for the underlying outer shell spectrum were fixed at the Shell values. Then we were able to obtain satisfactory fits for all of the inner regions ($\chi_{\nu}^2 \la 1.4$). It was required to fit O and Ne abundances in the ejecta model component for these inner radial regions. The overall O and Ne abundances for the ejecta component in these inner regions are significantly enhanced compared to the Shell by $\sim$3--6 times, which are consistent with those for Regions 1 and 2. Our best-fit elemental abundances among individual inner ejecta regions are consistent within uncertainties. The best-fit model parameters for these radial regions are listed in Table 2.

\subsection{Limits on the Compact Remnant}
Based on our deep {\it Chandra} data we place a tight upper limit on the embedded compact remnant that must have been created in this CC explosion. We performed MARX (http://space.mit.edu/ASC/MARX/) simulations to create point-like sources with a wide range of the 0.3--3 keV band luminosity assuming a power-law spectrum with a photon index of 2 and the foreground column appropriate for B0049--73.6. We added these simulated point sources at the center of the SNR's broadband image to run the point source detection. Then we estimate a 3$\sigma$ upper limit on the 2--7 keV band luminosity for the point source to be $L_{X}\sim$ 8.5 $\times$ $10^{31}$ erg s$^{-1}$.

\section{\label{sec:disc} DISCUSSION}
The central ejecta nebula is enriched in O, Ne, Mg, and Si. While O-, Ne-, and Mg-rich ejecta appear to show {\it smooth} distributions along the bright ring-like feature, our estimated Si abundances are significantly different between Regions 1 and 2. The overall metal abundance within Regions 1 and 2 are significantly higher than that of the Shell region which indicate their metal-rich ejecta origin. Region A's spectrum indicates the presence of ejecta material that is enhanced in Ne and Si. It is remarkable that Region A extends from within the central ejecta nebula out to the outermost ISM shell. The Ear is a faint feature and clearly extends beyond the outer swept-up shell. Our results from a single plane-shock model fit of this feature suggest a significantly enhanced O and Ne abundance (although uncertainties are large due to low photon statistics), O $=0.48^{+0.72}_{-0.29}$ and Ne $=0.96^{+0.79}_{-0.50}$ ($\chi_{\nu}^2 = 26/24$). We show its observed spectrum and best-fit model in Figure 3a. The suggested Ne overabundance in the Ear region is consistent with the bright Ne line EWs shown in Figures 2b $\&$ 2c. Based on multi-wavelengths source catalogs through the HEASARC on-line database, we found a red giant projected within the boundary of the Ear. Considering the extended X-ray emission feature with likely-enhanced metal abundances for the Ear, this red giant is unlikely the counterpart for the Ear, and thus we conclude that the Ear is associated with the SNR. These results suggest that Regions A and Ear may be high-velocity ejecta material, similar to ejecta bullets found in other SNRs (e.g. Vela shrapnels [Aschenbach et al. 1995, Miyata et al. 2001], an ejecta bullet in N49 [Park et al. 2012]). These overall ejecta features show that the spatial distribution of metal-rich stellar debris in B0049--73.6 is far from spherically-symmetric. We note that, while the metal-rich ejecta shows an asymmetric distribution, the outermost forward shock boundary of B0049--73.6 is nearly circular. Such differential morphologies between the ejecta and swept-up shell have been observed in other relatively aged ($\ga$ several $10^{3}$ year old) CC SNRs such as G292.0+1.8 (Park et al. 2007, Winkler et al. 2009) and Cygnus Loop (Katsuda et al. 2008).

Our radial spectral model fits show that overabundances for ejecta elements dominate the inner regions of $r\la 40 \farcs$. These radial structures suggest that the central metal-rich ejecta nebula extends roughly out to $r\sim40\farcs$, corresponding to $\sim$10 pc from the SNR center at a distance of 60 kpc to the SMC. We show the distributions of the broadband surface brightness and $\chi_{\nu}^2$ from our one-component shock model fits along the radius of B0049--73.6 in Figures 4a \& 4b. We also show the radial profile of the Ne abundance based on our best-fit spectral models (two-component model at $r\la$ 10 pc and one-shock at $r\ga$ 10 pc in Figure 4c). The O abundance shows a similar radial profile to that of the Ne abundance. For completeness we show the fractional flux contribution from the ejecta-like emission component based on our two-component model fits for all radial regional spectra along the radius of B0049--73.6 (Figure 4d). We confirm that the contribution from the ejecta-like emission component in the total observed flux dominates for regions $r\la$ 9 pc and is significantly smaller for regions $r\ga$ 9 pc from the SNR center. It is evident that all of these parameters show significantly higher values at $r\la$ 9 pc than those at $r\ga$ 9 pc from the SNR center. These suggest the location of the CD at $r\sim$ 9 pc, which extends significantly beyond the brightest part of the central ejecta nebula. This larger estimate for the CD is generally consistent with standard SNR dynamics as seen in one-dimensional hydrodynamical simulations by H05. A significant effect from the previously-suggested Fe-Ni bubble would not be required for our estimated size of the CD.

Based on the average values of the ejecta elemental abundances from Regions 1 and 2, we estimate the abundance ratios of O/Ne $=3.0^{+1.1}_{-1.0}$, O/Mg $=20.0^{+2.0}_{-1.4}$, and O/Si $=4.0^{+3.3}_{-3.0}$. These ratios are in agreement with H05 except for O/Si which we measured to be $\sim$3 times lower. We note that H05 did not measure the O abundance of B0049--73.6 and only assumed a value 0.2 in their abundance ratios. Also, our model fits use much better photon statistics and updated NEI modeling, and thus we consider our measurements more reliable than those by H05 (However, we note that our O/Si abundance estimate is based on two small regions where the Si abundance significantly varies. While we discuss the progenitor's nature based on these abundance ratios estimated from somewhat limited regions of the SNR, the abundance ratios measured from an extensive ejecta area [which is the subject of our follow-up work] would be required to reach a more conclusive interpretation). These abundance ratios are in plausible agreement with the CC SN nucleosynthesis models for a 13--15$M_{\odot}$ progenitor with solar or sub-solar ({\it Z} = 0.004) metallicity \citep{nomo06}. It is notable that the Shell is significantly under-abundant compared to the general SMC values (RD92) by a factor of $\sim$3--4 for O, Ne, Mg, and Fe. This suggests that B0049--73.6 might have exploded in a locally metal-poor environment, which may support a low-{\it Z} progenitor.

Based on the volume emission measure ({\it EM}) estimated from the spectral fit of the Shell we calculate the post-shock electron density ({\it n$_e$}). For this estimate we assumed the X-ray emitting volume of {\it V} $\sim3.0 \times 10^{58}~cm^3$ for our Shell region for a $\sim$3 pc path length (roughly corresponding to the angular thickness of the Shell at 60 kpc) along the line of sight. We assumed $n_e\sim1.2 n_H$ for a mean charge state with normal composition (where $n_H$ is the H number density). Then our estimated electron density is $n_e\sim2.4~f^{-\frac{1}{2}}~cm^{-3}$, where $f$ is the volume filling factor of the X-ray emitting gas. The pre-shock H density is  $n_0\sim0.5~f^{-\frac{1}{2}}~cm^{-3}$ assuming a strong adiabatic shock where $n_H = 4$ $n_0$. Based on their Sedov model fits H05 found that the post-shock regions of B0049--73.6 are in an electron--ion temperature equipartition. Ghavamian et al. (2007) found that several middle-aged SNRs (such as Cygnus Loop and DEM L71) have relatively low shock velocities and as such would be in electron--ion temperature equipartition. We thus assume electron--ion temperature equipartition for B0049--73.6 and then the gas temperature is related with the shock velocity ($v_s$) as $T=3\hat{m}v_s^2/16k$ (where $\hat{m}\sim0.6m_p$ and $m_p$ is the proton mass). For gas temperature of $kT=0.26$ keV for the outer swept-up shell, we calculate a shock velocity of $v_s$ $\sim$ 460 km s$^{-1}$. For the SNR radius of $\sim$20.5 pc we estimate the Sedov age of $\tau_{sed} \sim17200$ yr for B0049--73.6 (in the case that the electron--ion temperature equipartition has not been established yet in B0049--73.6, our calculated $v_s$ would be considered a lower limit, and therefor our SNR age estimate would be an upper limit.) We calculate the corresponding explosion energy of $E_0\sim1.7 \times 10^{51}$ erg. The total swept-up mass, $M_{SW}=n_0m_pV$, is estimated to be $M_{SW}\sim440~f^{\frac{1}{2}}~M_{\odot}$. Our estimated Sedov age for B0049--73.6 is consistent with that of H05 within $\sim$20$\%$. Our estimated $E_0$ is somewhat larger than the previous estimate (H05), but is consistent with the canonical value for a SN within a factor of $\sim$2. These results suggest that B0049--73.6 was probably created by a CC SN explosion (with a normal $E_0$) of a 13--15$M_{\odot}$ progenitor with a low metallicity.

Our estimated upper limit on the X-ray luminosity of an embedded compact stellar remnant at the center of  B0049--73.6 is more than 2 orders of magnitude lower than the previous estimate ({\it L} $\sim$ 1.5 $\times$ $10^{34}$ erg $s^{-1}$ [Weisskopf \& Hughes 2006]). Our tight upper limit on X-ray luminosity of the embedded compact stellar remnant suggests a low-luminosity source, probably an object belonging to the Dim Isolated Neutron Star class (e.g., Mereghetti 2011).

\acknowledgments

This work has been supported in part by {\it Chandra} grant GO0--11075A.

\clearpage

\begin{deluxetable*}{ccccccccccc}
\footnotesize
\tabletypesize{\scriptsize}
\setlength{\tabcolsep}{0.05in}
\tablecaption{Summary of Spectral Model Fits to Subregions of SNR B0049-73.6}
\label{tbl:tab1}
\tablewidth{0pt}
\tablehead{ \colhead{} & \colhead{$N_{\rm H,SMC}$} & \colhead{$kT$} & \colhead{$n_et$} & \colhead{\it EM} & \colhead{$\chi^2$/n} & \colhead{O} & \colhead{Ne} & \colhead{Mg} & \colhead{Si} & \colhead{Fe} \\
\colhead{Region} & \colhead{(10$^{21}$~cm$^{-2}$)} & \colhead{(keV)} & \colhead{(10$^{12}$ cm$^{-3}$ s)} & \colhead{($10^{57}$ cm$^{-3}$)} & \colhead{} & \colhead{} & \colhead{} & \colhead{} & \colhead{} & \colhead{}}
\startdata
Shell & $0.15^{+0.04}_{-0.05}$ & $0.26^{+0.04}_{-0.02}$ & $0.25^{+0.06}_{-0.06}$ & $80.50^{+30.90}_{-40.20}$ &  136/99 & $0.03^{+0.01}_{-0.01}$ & $0.04^{+0.01}_{-0.01}$ & $0.06^{+0.04}_{-0.03}$ & 0.30\tablenotemark{a} & $0.04^{+0.01}_{-0.01}$\\
A\tablenotemark{b} & 0.15\tablenotemark{c} & $0.40^{+0.04}_{-0.03}$ & $>5.00$ & $6.10^{+0.70}_{-1.40}$ 
& 125/88 & $0.07^{+0.03}_{-0.04}$ & $0.40^{+0.05}_{-0.08}$ & $0.14^{+0.05}_{-0.06}$ & $0.77^{+0.30}_{-0.25}$ & $0.04$\tablenotemark{d}\\
1 & 0.15\tablenotemark{c} & $0.32^{+0.03}_{-0.03}$ & $1.50^{+2.20}_{-0.48}$ & $30.10^{+3.00}_{-5.00}$ & 137/93 & $0.18^{+0.03}_{-0.03}$ & $0.34^{+0.05}_{-0.05}$ & $0.24^{+0.07}_{-0.07}$ & $1.16^{+0.37}_{-0.31}$ & $0.10^{+0.01}_{-0.03}$\\
2 & 0.15\tablenotemark{c} & $0.34^{+0.01}_{-0.03}$ & $1.10^{+1.30}_{-0.18}$ & $28.00^{+8.20}_{-2.50}$ & 158/92 & $0.12^{+0.01}_{-0.01}$ & $0.24^{+0.02}_{-0.03}$ & $0.15^{+0.02}_{-0.03}$ & $0.37^{+0.10}_{-0.07}$ & $0.04$\tablenotemark{d} \\
\enddata
\tablecomments{Abundances are with respect to solar \citep{ande89}. Uncertainties are with a 90\% confidence level. The Galactic column $N_{\rm H,Gal}$ is fixed at 0.45 $\times$ 10$^{21}$ cm$^{-2}$. For comparisons the RD92 values for SMC abundances are O = 0.126, Ne = 0.151, Mg = 0.251, Si = 0.302, Fe = 0.149.}
\tablenotetext{a}{Si abundance was fixed at RD92 value for the SMC.}
\tablenotetext{b}{The best-fit parameters of the ejecta component.}
\tablenotetext{c}{$N_{H,SMC}$ was fixed at best-fit Shell value.}
\tablenotetext{d}{Fe abundance was fixed at the best-fit Shell value.}
\end{deluxetable*}

\begin{deluxetable*}{cccccccc}
\footnotesize
\tabletypesize{\scriptsize}
\setlength{\tabcolsep}{0.05in}
\tablecaption{Summary of Spectral Model Fits to Radial Regions of SNR B0049-73.6}
\label{tbl:tab2}
\tablewidth{0pt}
\tablehead{ \colhead{Distance from} & \colhead{Counts} & \colhead{$kT$} & \colhead{$n_et$} & \colhead{\it EM} & \colhead{$\chi^2$/n} & \colhead{O} & \colhead{Ne} \\
\colhead{center (\farcs)}  & \colhead{} & \colhead{(keV)} & \colhead{(10$^{12}$ cm$^{-3}$ s)} & \colhead{($10^{57}$ cm$^{-3}$)} & \colhead{} & \colhead{} & \colhead{}}
\startdata
8\tablenotemark{a}  & 2400 & $0.35^{+0.09}_{-0.03}$ & $>1.00$ & $3.10^{+1.40}_{-1.30}$ &  59/48 & $0.15^{+0.10}_{-0.05}$ & $0.31^{+0.15}_{-0.09}$\\

20\tablenotemark{a} & 4100 & $0.35^{+0.05}_{-0.04}$ & $1.10^{+1.00}_{-0.46}$ & $5.80^{+1.70}_{-1.70}$ &  76/56 & $0.14^{+0.04}_{-0.04}$ & $0.30^{+0.07}_{-0.06}$\\

28\tablenotemark{a} & 4400 & $0.31^{+0.04}_{-0.03}$ & $>1.12$ & $10.50^{+1.90}_{-2.00}$ &  77/67 & $0.15^{+0.04}_{-0.03}$ & $0.26^{+0.05}_{-0.04}$\\

36\tablenotemark{a} & 3200 & $0.33^{+0.10}_{-0.03}$ & $>0.80$ & $3.60^{+1.60}_{-1.70}$ &  53/55 & $0.17^{+0.10}_{-0.09}$ & $0.30^{+0.14}_{-0.10}$\\

44 & 3100 & $0.40^{+0.08}_{-0.06}$ & $0.04^{+0.02}_{-0.01}$ & $9.80^{+3.70}_{-3.40}$ &  78/56 & $0.03^{+0.01}_{-0.01}$ & $0.04^{+0.01}_{-0.02}$\\

52 & 3600 & $0.36^{+0.06}_{-0.03}$ & $0.05^{+0.03}_{-0.02}$ & $3.80^{+1.00}_{-1.10}$ &  71/48 & $0.03^{+0.01}_{-0.01}$ & $0.07^{+0.02}_{-0.01}$ \\

60 & 3900 & $0.30^{+0.04}_{-0.02}$ & $0.15^{+0.10}_{-0.05}$ & $7.30^{+2.30}_{-2.60}$ &  80/55 & $0.03^{+0.01}_{-0.01}$ & $0.05^{+0.02}_{-0.01}$\\

68 & 3900 & $0.24^{+0.03}_{-0.02}$ & $0.18^{+0.13}_{-0.07}$ & $14.50^{+4.80}_{-6.10}$ &  48/53 & $0.03^{+0.01}_{-0.01}$ & $0.04^{+0.01}_{-0.01}$\\

\enddata
\tablecomments{Abundances are with respect to solar \citep{ande89}. Uncertainties are with a 90\% confidence level. The Galactic column $N_{\rm H,Gal}$ is fixed at 0.45 $\times$ 10$^{21}$ cm$^{-2}$ and the SMC column $N_{\rm H,SMC}$ is fixed at the Shell value 0.15 $\times$ 10$^{22}$ cm$^{-2}$. The Si abundance was fixed at RD92 value. The Mg and Fe abundance were fixed at the best-fit Shell value. The distance is from the SNR center to the center of each region corresponding to Figure 4(c).}
\tablenotetext{a}{For these regions the best-fit parameters of the ejecta component are presented.}
\end{deluxetable*}
\clearpage

\begin{figure*}[]
\figurenum{3}
\centerline{\includegraphics[angle=0,width=\textwidth]{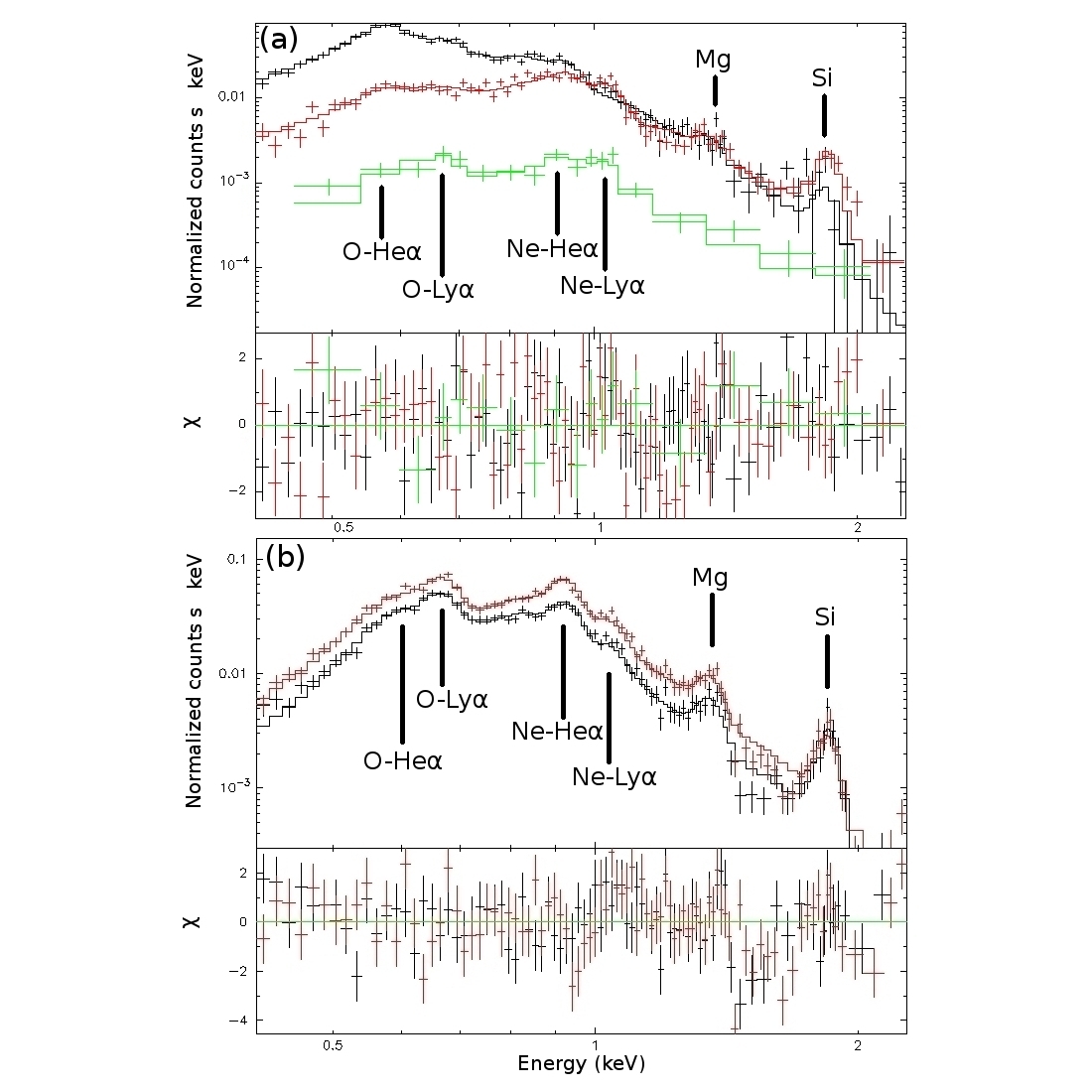}}
\figcaption[]{{The best-fit models and residuals of X-ray Spectra from selected regions within B0049--73.6 are shown. Several atomic emission line features are marked. (a) The ISM region is in black, Region A is in red, and the Ear region is in green. (b) Region 1 is in black and Region 2 is in red.}\label{fig:fig3}}
\end{figure*}

\begin{figure*}[]
\figurenum{4}
\centerline{\includegraphics[angle=0,width=\textwidth,height=0.8\textheight,keepaspectratio]{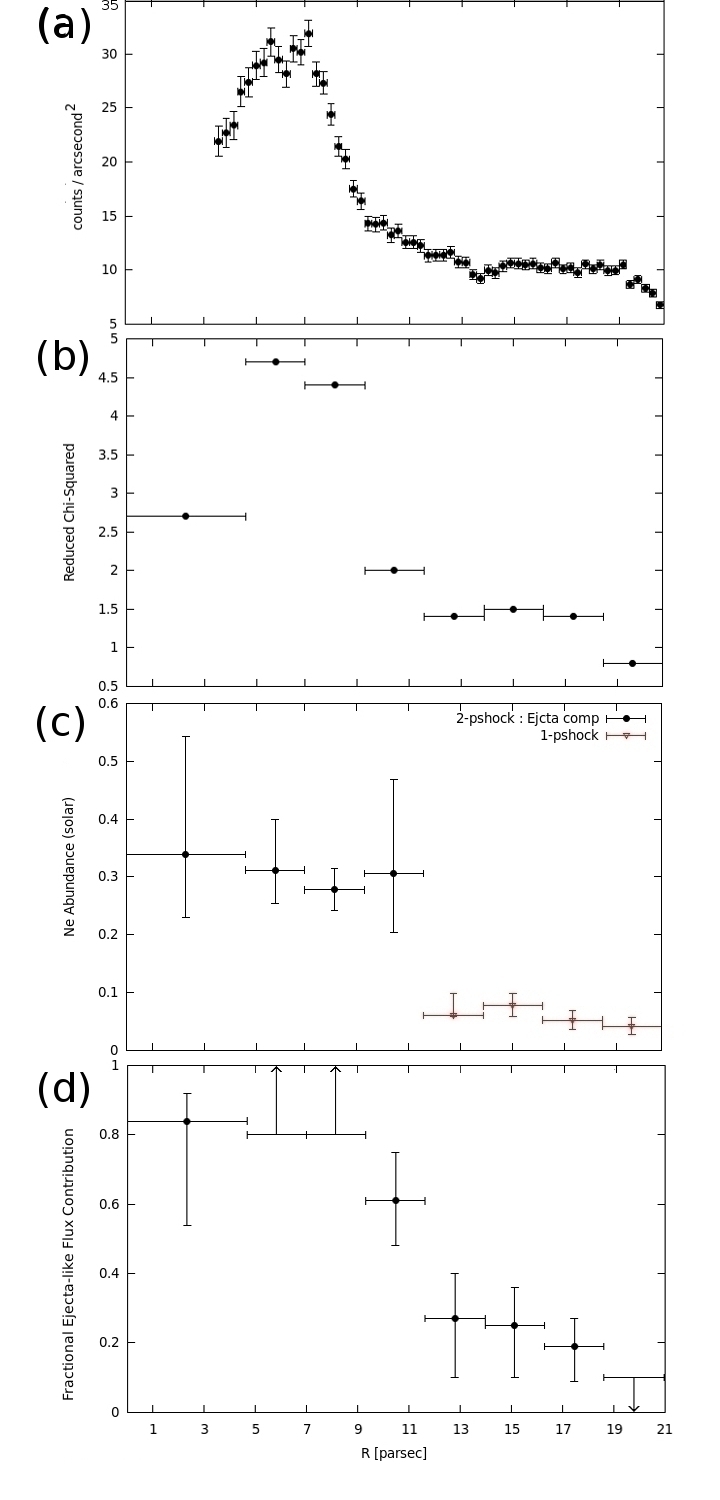}}
\figcaption[]{{(a) A radial profile of the broadband surface brightness for B0049--73.6. Each radial bin size is $\sim$0.3 pc ($1\farcs$). This radial profile is in the same direction as Region 2 (Figure 1b). 1$\sigma$ uncertainties are shown. (b) $\chi_{\nu}^2$ values from our one-component shock model fits with fixed Shell abundances along the radius of B0049--73.6. (c) The best-fit Ne abundance values along the radius of B0049--73.6. For $r\la$10 pc, they were based on the two-shock model fits (see text). (d) The fractional ejecta-like flux contribution based on our two-component model fits along the radius of B0049--73.6. Uncertainties for (c) and (d) are with a 90\% confidence level.}\label{fig:fig4}}
\end{figure*}

\end{document}